# Cellular Automaton for Realistic Modelling of Landslides

E. Segre
Istituto di Cosmo-Geofisica del C.N.R., Torino, Italy

C. Deangeli
Politecnico, Dipartimento di Georisorse e Territorio, Torino, Italy

**Abstract.** A numerical model is developed for the simulation of debris flow in landslides over a complex three dimensional topography. The model is based on a lattice, in which debris can be transferred among nearest neighbors according to established empirical relationships for granular flows. The model is then validated by comparing a simulation with reported field data. Our model is in fact a realistic elaboration of simpler "sandpile automata", which have in recent years been studied as supposedly paradigmatic of "self-organized criticality". Statistics and scaling properties of the simulation are examined, and show that the model has an intermittent behavior.

## 1  Background

Landslides are among the processes responsible for the evolution of the shape of the earths surface (Scheidegger, 1991). They constitute a major natural hazard, and therefore their prediction and modelling are very important in the environmental sciences.

The approach generally undertaken in soil engineering for the study of landslides concentrates on the phase of the slope failure, i.e., the analysis of the conditions for the breakdown of an elastic soil (see, e.g., Hutchinson, 1988). This approach, despite being fundamental for the processes referred in geotechnics as falls, topples, slides or spreads, which actually initiate a landslide, does not apply after the complete liquefaction of the soil mass, and does not describe the subsequent debris flow.

The establishment of constitutive equations for debris flow is, however, not trivial. Debris flow is in general described as a heterogeneous suspension of granular soil particles in water, in proportions and with grain size distributions that may largely vary. The overall motion is clearly too complex to be described by tracking each particle, and therefore the mixture is viewed as a continuum. The granular nature of the debris gives rise to a non-Newtonian fluid-like behavior. Since the seminal work of Bagnold (1954), a huge number of rheological models have been proposed in the literature for different possible regimes of these flows (see e.g. Chen, 1987; Sauret, 1987; Takahashi, 1991). Both the nomenclature and the models vary largely from author to author, and are often differentiated according to the initiating causes or the kind of soil involved in the slide (Pierson and Costa, 1987).

Since landslides have catastrophic effects, and cannot often be studied effectively under controlled conditions, one is led toward their computer simulation. To the best of our knowledge, computational debris flow models previously given in the literature have been quite sparse (Sassa, 1988; Vulliet and Hutter, 1988; Takahashi, 1991), and were based on finite elements implementations of the assumed fluid equations.

In contrast, in this paper we build a model which includes a stopping and re-initiation mechanism that can give rise to criticality. Our aim, apart from that of to simulate effectively landslides, is to reconnect with reported results for the so-called *cellular automata*.

Cellular automata are computational lattice models in which each reticular site can be in one of a limited number of states, whose evolution is solely determined by local rules. In recent years these models have been employed in many different physical computations. Some of them, later called *lattice gases*, have been proved to conform in the thermodynamic limit to important macroscopic differential equations (Frisch et al., 1986; Chen et al., 1992, 1993). In this respect cellular automata can be translated into algorithms which are significantly more efficient than corresponding finite differences integration schemes, and are straightforwardly applied to complex boundary geometries.

Moreover, a particular cellular automaton, namely the "sandpile model" of Bak et al. (1987), has been introduced in connection with avalanche dynamics. That automaton has been used as a toy model to explain the occurrence of power-law statistics in natural processes, via a supposedly universal feature which has been named *self-organized criticality*. Since the original paper, there has been much discussion in the literature about the real significance of that model, and about its applicability to real granular flows. Various elaborations of the original automaton of Bak et al. (1987) have been reported since then. Some of these elaborations aim to achieve more realistic effects, and generate different statistical and scaling properties. Recent contributions to the subject include for instance those by Dhar (1990); Prado and Olami (1992); Ding et al. (1992); Christensen et al. (1992); Krug (1993); Theiler (1993); Rosendahl, et al. (1993); Carlson, et al. (1993); Morales-Gamboa et al. (1993); Chhabra et al. (1993); Maslov and Olami (1993); Frette (1993). Experiments with real sand grains, as well with other granular materials have been carried out for example by Nagel (1992); Evesque et al. (1993); Nori and Pla (1993), and compared with the expectations from theoretical sandpile automata. A comprehensive review of the experimental and computational investigation on granular -and specifically sand- flows can be found in Metha and Barker (1994). A descriptive contribution by Noever (1993), which considers statistics of landslides in a mountain region, is also to be mentioned in this context.

The aim of this paper is to propose an effective cellular model for debris flow, combining the schematicity of sandpile models and more appropriate geotechnical empirical laws, and to look at its statistical behavior. To our knowledge, a similar approach has not yet been undertaken for debris flow in any other place than the note of Digregorio et al. (1994), which appeared after the first version of this paper was submitted. However, a much simpler cellular automaton is considered there, and the attention is mainly posed on its parallelization.

The paper is organized as follows: the rules of our automaton are derived from established debris flow laws in Sect. 2; the scheme is numerically tested in Sect. 3; a real event is simulated in order to calibrate the model in Sect. 4, and finally a multifractal analysis of the resulting time series is reported in Sect. 5.

## 2 Computational model

### 2.1 Lattice setup

We are primarily interested in describing the flow of a heterogeneous and scarcely coherent debris, such as that of a stony discharge, which takes part over a rigid and non erodible substratum. We furthermore assume fully mixed debris (*mature debris flow* in the nomenclature of Takahashi, 1991), i.e., a homogeneously saturated debris layer. The appearance of surface flowing water, the shear stability of different strata, etc., will not be upon question here. Our model does not account for varia-tions of the debris properties in the vertical direction $z$, i.e., we adopt a vertically averaged description. This allows us to schematize to some extent the real dynamics.

**Table 1.** Data values interpolated for the constitutive relations

| $s_1$ (silt) | $s_2$ (gravel) | $s_3$ (boulders) | $\rho_s$ $g/cm^3$ | $\phi$ degrees |
|---|---|---|---|---|
| 100% | 0% | 0% | 1.8 | 12° |
| 0% | 100% | 0% | 2.0 | 30° |
| 0% | 0% | 100% | 2.2 | 35° |

The landslide site is discretized in elementary cells of finite size. In each one the state of the system is specified by the values of some representative quantities. These include: the height of the impermeable rigid substratum $r$; the nonnegative amounts of water $w$, of gas (or void fraction) $g$ and of granular solids $s$ in the cell. In order to account for debris properties that depend on the granulometry, the solid content is classified by size. Three size classes have been presently employed, representing silt-clay, sand-gravel and boulders. All the contents are given as partial heights (volumes/base area of the cell), so that the top height in the cell $i$ is given by

$$z(i) = r(i) + w(i) + g(i) + \sum_{k=1}^{3} s_k(i), \quad (1)$$

and the volume of its content is $V = S[z(i)-r(i)]$, where S is the base area of the cell.

The density of the solid part $\rho_s$ and the "friction angle" $\phi$ depend on the composition of the debris. In first approximation they are calculated by linear interpolation of the data points given in Table 1, as

$$\rho_s(i) = \frac{\sum_{k=1}^{3} \rho_k s_k(i)}{\sum_{k=1}^{3} s_k(i)}, \; \phi(i) = \frac{\sum_{k=1}^{3} \phi_k s_k(i)}{\sum_{k=1}^{3} s_k(i)}, \quad (2)$$

i.e., the two quantities are expressed as weighted means of the ones of the individual solid components.

More sophisticated empirical laws, detailing different soil classes, could be employed in a future refinement; the form of $\rho_s$ used here allows to impose the conservation of mass simply by conserving the volume of the transported material, and fits satisfactorily to reported soil data.

The embedding fluid is assumed to be clear water, with density $\rho_f = 1 g/cm^3$; the weight of gas is neglected. The solid concentration $C$ and the density $\rho$ of the mixture in cell $i$ are thus

$$C(i) = \frac{\sum_{k=1}^{3} s_k(i)}{g(i) + w(i) + \sum_{k=1}^{3} s_k(i)}, \quad (3)$$

$$\rho(i) = C(i)\rho_s(i) + (1 - C(i))\rho_f. \quad (4)$$

Each cell is considered connected with a number $b$ of its nearest neighbors. At each discrete time step some material can be transferred to those neighbors, according to rules which take into account the local slope in the neighbors directions. The whole lattice is thus viewed as a network of elementary slopelets, each one with its characteristic slope due to the differential topography and debris accumulation.

**Fig. 1.** Different lattice geometries based on a cartesian mesh: a) square, b) hexagonal, c) octagonal, d) triangular, e) 16-stellar. Note that in case b) the white points constitute a shifted independent lattice.

Different lattice geometries, corresponding to different choices of meshes and sets of $b$ first neighbors, can be devised. The most obvious arrangements are the Cartesian square lattice ($b = 4$) and the hexagonal lattice ($b = 6$). Other staggered stencils are easily implemented (Fig. 1). A geometrical realization of the top surface may have to be dropped, because elementary slopelets cross themselves as the stencil is shifted. Such arrangements may be, however, effective in smoothing potential discretization instabilities (see below in Sect. 3); the set of local slopes is then interpreted as a pure property assigned to the lattice. In such cases the fact that cell $j$ is a neighbor of cell $i$ may not even imply the converse; consider for instance what is obtained by shifting the pattern of Fig. 1d along the lattice.

We have to resort to a two-neighbors partition rule for the transfers, in order to achieve lattice isotropy (Fig. 2). By this we mean that the flow along an arbitrarily inclined plane must not depend on the orientation of the lattice. To this extent a local slope angle $\theta(i, v)$ is defined in the cell $i$ for each *pair* of its adjacent neighbors $\{j_v, j_{v+1}\}$ ($v = 1, \ldots, b$ modulo $b$). The steepest descent vector $\vec{p}(i, v)$ on the plane passing for the three center points is

$$\vec{p}(i, v) = (\vec{j}_v, \Delta z_v(i)) \wedge (\vec{j}_{v+1}, \Delta z_{v+1}(i)), \qquad (5)$$

where $\vec{j}_v$ denotes the two-dimensional lattice vector oriented from node $i$ to $j_v$ and $\Delta z_v(i) = z(j_v) - z(i)$. The slope angle is then

$$\theta(i, v) = \tan^{-1} \frac{|\vec{p}_{xy}(i, v)|}{-\vec{p}(i, v) \cdot \vec{z}}, \qquad (6)$$

where $\vec{z}$ is the unit versor in the $z$ direction, and $\vec{p}_{xy}(i, v)$ the projection of $\vec{p}(i, v)$ onto the $xy$ plane. The outflow rate $\vec{q}(i, v)$ from the cell $i$ toward the sector $\{j_v, j_{v+1}\}$ is assumed to be parallel to $\vec{p}(i, v)$ and function of $\theta(i, v)$. For this purpose $\vec{p}_{xy}(i, v)$ is decomposed as

$$\vec{p}_{xy}(i, v) = p_1(i, v)\vec{e}_v + p_2(i, v)\vec{e}_{v+1}, \qquad (7)$$

$$\begin{aligned} p_1(i, v) &= \frac{\vec{p}(i, v)}{|\vec{p}(i, v)|} \cdot \frac{\vec{e}_v - (\vec{e}_v \cdot \vec{e}_{v+1})\vec{e}_{v+1}}{1 - (\vec{e}_v \cdot \vec{e}_{v+1})^2}, \\ p_2(i, v) &= \frac{\vec{p}(i, v)}{|\vec{p}(i, v)|} \cdot \frac{\vec{e}_{v+1} - (\vec{e}_v \cdot \vec{e}_{v+1})\vec{e}_v}{1 - (\vec{e}_v \cdot \vec{e}_{v+1})^2}, \end{aligned} \qquad (8)$$

where $\vec{e}_v = \vec{j}_v / |\vec{j}_v|$ is the $v$-th lattice versor.

We impose volume and, consequently, mass conservation separately for each solid class and for water. The gas balance will be taken into account differently, since free exchange can take part with the atmosphere. A more elaborate model could eventually consider either fragmentation processes, in which some amount of the larger granular classes may break and add up to the smaller ones. Alternatively, there may be more overlaid strata, each one with its composition, with possible mutual exchanges. In any event, we obtain a model which has, in the sandpile terminology, local and limited rules.

Energy and momentum conservation are not enforced here. This is consistent with modelling a process which is dissipative in character, both because of the increasing dispersion and liquefaction and because of viscosity. The assumption that the debris previously in motion can suddenly be stopped within the space of a single cell, depending only on instantaneous local conditions (as modelled below), is equivalent to the assumption that kinetic energy is readily dissipated, gravity being the main energy source.

Initial conditions for the model are imposed specifying the site topography and the debris distribution at the initial time. Since our primary aim is to model debris discharges, which possess typical time scales ranging from seconds to minutes, we will consider the total water amount as fixed. Even if the primary cause which triggers landslides is usually an extreme rainfall, the water content of the soil varies on a longer time scale, and need not to be externally varied during the simulation. For the same reason, we think that permeation through the soil has no time to occur in this regime of flow, and we do not consider the use of conductivity laws, such as Darcy's law. Net in- or out-fluxes in particular source or pit cells can, however, readily be included into the model, and could be required as boundary conditions. This would be the case, for example, if the computational domain had to be truncated up or downstream, or if permeation through the substratum intervened significantly into the water balance.

**Fig. 2.** Schematic of the debris transport process (depicted for square cells, but easily transposed to other geometries).

Boundary conditions are easily implemented on any set of cells: to realize an open boundary it is just sufficient to force the content of these cells to be always null, while to achieve closed boundaries one simply sets to zero any resulting out-flux. Neighbors relative positions could be calculated modulo the lattice dimensions to achieve periodic boundary conditions, if ever necessary.

## 2.2 Deriving automaton rules from empirical flow laws

Evolution rules for the automaton are imposed consistently with established empirical laws for debris flow. These rules must account for the initiation ("toppling" in the sandpile literature), motion and stopping of the debris mixture. As a general issue (see Chen, 1987), we note that often the rheological models underlying the empirical laws may not be completely correct. The model equations may not be able to account for the observed values when microscopic quantities (such as the actual grain sizes) are substituted. The proposed formulas can nevertheless sometimes fit real data, if one relaxes the assumptions that were made to derive them or introduces ad hoc empirical factors. This fact encourages us to make use of rather empirical rules, regarding the approximate agreement with real data as the main validation.

Under the conditions exposed at the beginning of the previous paragraph, the debris flow is in the so called *granulo-inertial* range (Seminara and Tubino, 1990; Takahashi, 1991). In this dynamical range the forces due to the turbulent motion of the embedding fluid are negligible with respect to the collisional stresses of the solid fraction. The macroscopic behavior of the granular mixture is assimilable to that of a dilatant viscous fluid, i.e., a non-Newtonian fluid with a quadratic dependence of the viscous forces upon the shear stresses.

We model the elementary slip processes solely with a state of rest and a state of flow. A rationale for doing this is that the distance required for complete liquefaction of a debris mass is short (see Takahashi, 1991, Figure 3.6). Therefore, in a simplified model, we neglect the details of the (slope failure) initiation phase of the movement. We will also make use of the assumption, perhaps strong, but adequate for our purposes, that the elementary flows along the lattice take the form of granulo-inertial stationary debris flow in uniform rigid bottom channels.

### 2.2.1 Initiation

A condition for the occurrence of debris flow on a slope plain is given by Chen (1987), Eq. (42): steady equilibrium flow takes part if the slope angle $\theta$ satisfies

$$\frac{\rho - \rho_f}{\rho} \sin \phi \leq \tan \theta \leq -\frac{\rho - \rho_f}{\rho} \frac{\mu_1}{\mu_2}. \tag{9}$$

In Eq. (9), $\phi$ is called the "static friction angle" and $\mu_1$ and $\mu_2$ represent respectively a "consistency" and a "cross consistency" coefficient of the squared shears in the expression of the dilatant viscous stresses. Various forms for the dependence of these coefficients on the concentration and on the granulometry are proposed in the literature, leading to different results and leaving much space to ad hoc fits. For example these coefficients are estimated by Chen (1987) to depend on the concentration as $A(1 - C/C_*)^{-2.5C_*}$, as is suggested by Einstein's equations for granular viscosity, with $C_*$ equal to the maximum admissible concentration and $A$ a constant. By referring to the original Bagnold's work (Bagnold, 1954), Takahashi (1991) instead expresses $\mu_1 = a\rho_s(\lambda d)^2 \sin \phi_d$, with $a$ an empirical constant, $d$ the solid particle diameter, $\lambda$ the linear concentration and $\phi_d$ the "Bagnold quasistatic stress angle".

Some disagreement also exists in the literature about the use in Eq. (9) of $\phi$ rather than $\phi_d$, which are found experimentally to differ by a few degrees. The lower bound on $\theta$ coincides with the Bagnold estimate if it is assumed that $\tan \phi_d = \sin \phi$ in the quasistatic limit. If, furthermore, $\mu_2 = -a\rho_s(\lambda d)^2 \cos \phi_d$ is assumed, Eq. (9) would tell that equilibrium steady uniform flow occurs only for $\theta = \phi_d$; that would however lead to other inconsistencies, as pointed out both by Chen (1987) and Takahashi (1991). Since in our study the angle $\phi$ is employed rather heuristically, we will not need to be more precise. More care would be needed if the appropriate value was to be compared with specific in situ measurements.

In our lattice model we neglect the upper bound and simply translate Eq. (9) into the rule: the flow initiates on every lattice sector $(i, v)$ for which

$$\tan \theta(i, v) > \frac{\rho(i) - \rho_f}{\rho(i)} \tan \phi(i). \tag{10}$$

The intervening $\rho(i)$, $\theta(i, v)$ and $\phi(i)$ are calculated in each cell as explained in Sect. 2.1. For every cell of the lattice, all the branches to neighbors $j_v$ such that Eq. (10) is satisfied will be called *critical directions*, and flow is assumed to move material to those neighbors only.

### 2.2.2 Flow

In the stationary flow of a granulo-inertial debris layer over a rigid bottom, the flow velocity $u$ has a vertical dependence on $z$ like (Chen, 1987, Eqs. 46–49):

$$u(z) = \frac{2}{3}\sqrt{\frac{\rho_* g \sin\theta}{\mu_1}}\left[z_0^{3/2} - (z_0 - z)^{3/2}\right], \tag{11}$$

where $g$ is the gravity acceleration, $\rho_* = \gamma\rho$ is a corrected density, with $0 < \gamma < 1$. In Eq. (11), $z_0$ is the height from the substratum at which the flow velocity becomes independent from $z$, e.g., where the shear stress equals the vertical component of the weight of the upper layer. For flows of incohesive granules, $z_0 = h$ (the whole layer of debris moves according to the law) and $\gamma \to 1$. Vertical averaging of Eq. (11) gives then the cross sectional velocity $U$ of the entire layer (Takahashi, 1991, Eq. 2.3.4):

$$U = \frac{2}{5}\sqrt{\frac{g\rho h^3 \sin\theta}{\mu_1}}. \tag{12}$$

This formula is consistent with dimensional estimates based on the classical Manning or Chezy forms of the stresses (Trowbridge, 1987; Scheidegger, 1991), which account for the presence of the square root of $\rho \sin\theta$.

If the concentration of the solids or the granulometric distribution are allowed to vary within the debris depth, as it is frequently observed, then the description becomes more complicate. A number of peculiar phenomena are observed in non-homogeneous granular flows; for instance the larger boulders tend to float over the bulk of the debris, can be transported farther apart and accumulate along the fronts. Qualitative explanations for the observed levitation of the larger grains are found in Savage and Lun (1988) and Takahashi (1991), §4.4.

As for the concentration of the flowing debris, a hint could be taken from the equilibrium concentration for the steady uniform inertial debris flow, which is

$$C = \frac{\rho_f}{\rho_s - \rho_f}\frac{\tan\theta}{\tan\phi - \tan\theta} \tag{13}$$

(Chen, 1987, Eq. 62; Takahashi,1991, Eq. 2.3.9, though inconsistent for very steep channels). The use of such an equation would imply to assume that exactly as much water takes part in the motion as is required to maintain the steady uniform debris flow. Since it is not assured that the required amount of water is actually available in the cell, we prefer to treat the different components as if they were moving independently. At the same time we can mimic the faster velocity of the larger grains.

We therefore implement a variant of Eq. (12) on the lattice. We let quantities $q$ of material flow out of the cell $i$ toward its critical neighbors and evaluate $q$ by putting $z(i) - r(i)$ as $h$ and $\theta(i,v)$ as $\theta$ in Eq. (12). We assume that the concentration and the density of the flow are those of the source cell $i$. The slope $\theta(i,v)$ would also decrease during the elementary motion. We would have to account for a temporal dependence of $h(t)$, obtained by solving the differential equation

$$\frac{dh(t)}{dt} = \frac{q(C,t)}{S} = \frac{2P}{5S}\sqrt{\frac{g\rho(t)h(t)^5 \sin\theta(i,v,t)}{\mu_1(t)}}, \tag{14}$$

in which $P$ is an appropriate effective width (hydraulic radius), and also the concentration $C(t)$, the density $\rho(t)$ and the slope inclination depend on time. In order to keep the model as simple as possible, we assume that the $q(C,\Delta z)$ are constant during the time step, and rely on the first order approximation for small values of the time step $\Delta t$.

Moreover, we compute independent $q_k$ for each granular fraction ($k = 1, 2, 3$) and for water ($k = 0$), and assume complete mixing before and after the transfer. The simplified evaluation is thus

$$q_k(i,v) = Q_k \Delta t \sqrt{\rho h^5 \sin\theta(i,v)}, \tag{15}$$

with $Q_k$ constants, conglobating and approximating all the neglected factors of Eq. (14). The ratios of the $Q_k$ entirely account for the faster propagation of the larger grains. Water is treated as a granular class, since we put $w(i) = s_0(i)$ and use Eq. (15) with a rate constant $Q_0$ is of the same order of the $Q_k$. However, water is differentiated from the solids since it plays a different role in the evaluation of $\rho$, of $\phi$ and in Eq. (10).

As mentioned before, the outflow from each sector $(i,v)$ of the cell reparts among two adjacent neighbors proportionally to $p_1$ and $p_2$ computed from Eq. (8). The time advancement rule will be

$$\begin{aligned}
s_k(i) &\to s_k(i) - [p_1(i,v) + p_2(i,v)]q_k(i,v) \\
s_k(j_v) &\to s_k(j_v) + p_1(i,v)q_k(i,v) \\
s_k(j_{v+1}) &\to s_k(j_{v+1}) + p_2(i,v)q_k(i,v),
\end{aligned} \tag{16}$$

where it is intended that $q_k(i,v) = 0$ if the $v$-th sector is not critical. Upper limits on $q_k(i,v)$ prevent the cells to be more than depleted within the time step: after all $q_k(i,v)$ have been calculated for the cell $i$,

$$q_k(i,v) \to q_k(i,v)\min\left(1, \frac{s_k(i,v)}{\sum_v q_k(i,v)}\right), \tag{17}$$

and only then the $s_k(i)$ are updated. The rule Eq. (16) with the correction of Eq. (17) is iterated over all $i$ and all $v$. All the computed $q_k(i,v)$ are stored before the simultaneous updating of the lattice. In this way the computation may also be completely parallelized.

We treat the void fraction in a similar way. Void, which has to be taken into account for the total volume, will in our simplification be class $-1$, moving around the lattice according to Eq. (15) with its own rate factor $Q_{-1}$. However, void is not conserved: experience shows that shaken granular material tends to unpack and hence to augment its volume (see e.g. Metha and Barker, 1994). To keep the model schematic, we let the running material increase proportionally its void ratio,

**Table 2.** Maximum allowed void ratio for each class

|       | $s_1$ | $s_2$ | $s_3$ |
|-------|-------|-------|-------|
| $e_{max}$ | 1.5 | 0.835 | 0.3 |

till a maximum value $e_{max}$. As a consequence, the void is propagated to the neighbors with the rule

$$s_{-1}(i) \to s_{-1}(i) - [p_1 + p_2] q_{-1}(i,v) \quad (18)$$

$$s_{-1}(j_v) \to s_{-1}(j_v) + p_1 \left[ q_{-1} + \alpha \sum_{k=0}^{3} q_k \right] (i,v)$$

$$s_{-1}(j_{v+1}) \to s_{-1}(j_{v+1}) + p_2 \left[ q_{-1} + \alpha \sum_{k=0}^{3} q_k \right] (i,v),$$

where $\alpha$ is a parameter to be calibrated, and the $q_k(i,v)$ have been calculated previously with Eqs. (15)–(17). Subsequently the upper limitation is enforced:

$$s_{-1}(i) \to \min\left( s_{-1}(i), e_{max}(i) \sum_{k=1}^{3} s_k(i) \right). \quad (19)$$

The limitation involves the maximum void ratio $e_{max}(i)$ admitted in the destination cell after the transfer. This ratio is calculated as that of a composite soil of corresponding composition, that is, by weighted average of the values given in Table 2 (Lancellotta, 1987, p. 7). The maximum void ratio is computed according to the solid fraction only.

The time advancement proceeds by discrete steps. The physical value of $\Delta t$ will be a consequence of the dimensional change that is allowed to take part in the system during a single step. This is in turn determined by the $Q_k$, which should be small enough to ensure the near constancy of $h(i)$ within a time step. An estimate of the temporal evolution of the outflow from a single cell is obtained by numerical integration of Eq. (14) and (15) as $\Delta t \to dt$. In such case the numerical integration shows an asymptotic relaxation of $h(t) \propto t^{-2/3}$.

Instantaneous flow rates $\vec{q}(i)$ are evaluated in each cell by vector summing all the incoming flows, that is,

$$\vec{q}(i) = \sum_{j'_v} \left[ q_{-1}(j'_v, i) + (1 + \alpha) \sum_{k=0}^{3} q_k(j'_v, i) \right] \vec{e}'_v, \quad (20)$$

where $j'_v$ labels all the cells for which $i$ is a first neighbor. Local instantaneous velocities are derived from these flow rates by dividing for $\Delta t$ and appropriate geometrical width factors.

No further assumptions are made about the dynamics of the elementary movement. The cell forgets the details, and smoothes the dishomogeneities after the advancement step has been made. The elementary transfer is not resolved below the $\Delta t$ scale.

We note that despite our scheme does not directly account for forces, it is not completely non dynamical, since it employs flow laws that have been derived with an underlying rheological model, and it still retains features of the viscous flow.

### 2.2.3 Stoppage or continuation

Very few quantitative arguments are given in the literature for the exact calculation of the stoppage conditions of a debris flow which runs on an undercritical slope. At least to our knowledge, the exact determination of the stoppage distance of a front is still an open problem in the general case (Takahashi, 1991, § 5.1). What it is at least established (Jaeger and Nagel, 1992; Metha and Barker, 1994) is that the repose angle at which collapsing dry granular material settles is by some degrees lower than the critical initiation angle.

We included in our model a simple local rule that accounts for the dynamical lowering of that angle. For this, a one-step memory is added to the model. If the cell received material at the previous time step, then still enough kinetic energy may be available to sustain the flow in a non critical direction. In this way a fast enough flow can even climb counter-slopes, as it is sometimes observed. The slope direction vector $\vec{p}(i,v)$ appearing in Eq. (6) is dynamically corrected in

$$\vec{p}(i,v) \to \vec{p}(i,v) + \beta \vec{q}(i, t - \Delta t), \quad (21)$$

where $\vec{q}(i, t - \Delta t)$ is the volume inflow in cell $i$ at the preceding time step, and $\beta$ is another parameter of the model. The flow then continues if the corrected $\theta(i,v)$ satisfies Eq. (10). In other words, it is assumed that an "effective slope" is seen by the running debris. The argument could be related to energy line considerations (see e.g. Sassa, 1988). A more elaborate model could take a different threshold into account, or allow for a partial deposition of material.

### 2.3 Summary of the computational model

For the algorithmic implementation of Eqs. (16), (17), (18) and (20), it results convenient to organize the variable information into two arrays. The first one is $s_k(i)$ ($-1 \leq k \leq 3$), and the second is the transfer array

$$G_k(i) = s_k(i, t + \Delta t) - s_k(i, t), \quad (22)$$

in which the differences result from the above mentioned equations. In other words, it is more convenient to track the in-fluxes coming from all $j'_v$ toward the cell $i$, than to track the out-fluxes going to each $j_v$.

The scheme requires, at each time step, the following operations to be performed:

- the state of each cell $i$ is defined by $r(i)$, $s_k(i)$, and by $\vec{q}(i)$ at the preceding step;

- $z(i)$, $\rho(i)$ and $\phi(i)$ are computed by Eqs. (1), (2–4);

  - $\theta(i,v)$ is computed by Eqs. (6) and (21) for all sectors of $i$;

- the critical directions are found by Eq. (10);
- the $q_k(i,v)$ are calculated by to Eq. (15–16);
- thresholds are enforced according to Eq. (17);
- the void increase is computed by Eq. (18);

- all gains and losses are included in $G_k(i)$;
- the boundary cells conditions are enforced on $G_k(i)$;
- the influx rates $\vec{q}(i)$ are computed by Eq. (20);
- the whole lattice is updated adding $G_k(i)$ to $s_k(i)$;
- the void limitation, Eq. (19), is enforced.

## 3 Testing the numerical model

### 3.1 Numerical stability

The implementation of the automaton with continuous state variables renders the explicit time advancement scheme only conditionally stable. We have observed, for example, the appearance of an oscillatory instability pattern. This can occur whenever, at a given time step, an exceeding quantity of material is transferred from a cell to its neighbor(s). If the material is then not drained away rapidly enough, then at the following time step the slope between these cells may result reversed, so that the the first outflow just returns into the original cell. If the mechanism is allowed to repeat, the back and forth transfer of matter gives rise to persistent oscillations, in which odd and even cells are alternatively depleted and filled. Averages either over a few time-steps or over a few cells remain well behaved in presence of this instability; but the unphysical oscillatory background velocities may amplify if $\beta$ is large enough. With our set of rules, enforcing ulterior thresholds to prevent this instability would be cumbersome. The best cure to this problem is a refinement of the time step (i.e., a rescaling of all $Q_k$), which scales down the elementary outflows.

The actual criterion for stability can be outlined in the following way: consider cell 1 and 2 upon a flat bottom, spatially separated by $l$, which contain at time $t$ respectively $h_1$ and $h_2$ heights of equal debris. Let the resulting slope angle be such that

$$\tan\theta_1 = \frac{h_1 - h_2}{l} > \tan\phi. \tag{23}$$

A partial height

$$d\Delta t \sim Q(1+\alpha)\Delta t\sqrt{\rho h_1^5 \sin\theta_1} \tag{24}$$

is then transferred from 1 to 2, where $Q$ is an average of the $Q_k$. Thus at time $t+\Delta t$ the effective slope, including the advection contribution, becomes

$$\tan\theta_2' = \tan\theta_1 - \frac{2d\Delta t}{l} + \frac{\beta d}{l^2}. \tag{25}$$

It must result $\tan\theta_2' > -\tan\phi$ in order to prevent a return of debris from 2 to 1, and therefore

$$\Delta t < \frac{l(\tan\theta_1 + \tan\phi)}{2d} + \frac{\beta}{2l}. \tag{26}$$

An upper bound on $\Delta t$ in terms of typical $h$, $l$, $Q$, $\rho$ can thus be derived. Since the outflow from one cell may be shared among more neighbors, this bound may be lowered by a factor up to $b$. In this sense lattices with higher coordination number may be numerically more stable.

### 3.2 Isotropy

In order to claim that our model faithfully reproduces the physical process, we must ensure that it has an isotropic behavior, at least in an averaged sense. This implies that the flow computation must result as independent as possible from the choice of the lattice geometry and from the orientation of the lattice with respect to the landscape features.

The two-neighbors transfer rule introduced in Sect. 2.1 guarantees a local spatial isotropy. We note that it is the use of flow laws such as those described in the previous sections, that renders necessary the two-branches mechanism. It can be easily checked that if a simpler single branch rule was employed, then the model would have been isotropic for any lattice geometry only if $q \propto \Delta z$, and no threshold on the minimal slope angle was enforced.

A theoretical verification of the isotropy requirement is not easy, because the flow evolution is determined by the continuous buildup of the landscape. It is not easy, for instance, to verify whether the flow reaches at the same time equally distant cells, regardless of their positions on the lattice. A detailed calculation of the transport between two sites would require to consider the contributions through every possible path connecting them. None of these paths is independent, because the material that accumulates along the paths alters the global landscape. The evaluation of the expectation or of the correlation values of the flux in different points or directions would be similarly affected. The arguments normally quoted for the isotropic behavior of lattice gases onto the hexagonal lattice (referring, e.g., to Frisch et al., 1986) do not apply to the present case, for the much too different setting of the problem.

Some isotropy in a time averaged sense is however numerically seen to be recovered. We choose to verify the isotropy of the model by considering the transport of matter in arbitrary directions. A number of numerical experiments can be carried out to this extent. We report two of them that convince us of the isotropic behavior.

To begin, it is verified that the flow rate is independent from the slope line orientation. Figure 3a shows the flow vectors obtained in a test simulation of the discharge of an uniform debris layer over a conical pedestal.

**Fig. 3.** flow vectors of an initially uniform distribution over a conical pedestal, after the first time step; b) scatter-plot of the velocity vectors of case (a) (the few off-circumference points represent either boundary cell rates, or are discretization artifacts)

**Fig. 4.** Accumulation conoid produced by a point source on a flat topography, after 40000 time steps of 0.02 sec with $\beta = 0$.

**Fig. 5.** Roll wave pattern in a numerical simulation of a channel debris flow with constant upstream alimentation. Grid resolution is $61 \times 61$.

The resulting flow is expectedly radially symmetric, and the flow vectors show, apart from effects due to the discretization, a radial distribution, as can be seen from the scatter-plot in Fig. 3b.

Another test, inspired by Prigozhin (1993), is to check that material emitted by a single source cell over a flat bottom accumulates in a circular based conoid. This is also achieved, and shown in Fig. 4. The parameters used in these two tests are the same used in the case simulation described below in Sect. 4.

3.3 Roll waves

Channelized steady debris flow can develop an undulatory free surface instability, namely the appearance of roll waves, which are often observed in real events. Conditions for the development of such instability, which imply an evaluation of the shear and of the bottom stresses of the debris, have been examined by Trowbridge (1987) for the Newton and for the Bingham fluid model. We were able to achieve a similar-looking pattern in the numerical simulation of a constant slope channel, with upstream constant alimentation and downstream open boundary conditions. The debris layer thickness after a long simulation time is presented in Fig. 5; here again the same parameters of the case studied in Sect. 4 have been used.

4 Case study of a real landslide: validation and back analysis of the model

The agreement of the computational result with a real, full size landslide validates our model. Since during the time of this work we had no direct access to a fully mon-itored event, we had to resort to a case history found in the literature (Jiang Yun et al., 1991). This event was chosen among others because of the extended interested region and the relative variety of phenomena, which included also counter-slope flows. We were able to reproduce the major features of this event from the data presented in that report, even conjecturing some of the missing information.

The landslide detached from the Xi Kou Mountain in the Sichuan province of China, on July 10, 1989. The slope where the landslide took place was composed by very weathered and fractured strata of limestone, marlite, dolomite and shale. The bedding had a dip opposite to that of the slope. The ridge of the slope was $1012m$ high, while the Xi Kou town is located on a depression at a height of $298m$. Between the heights of $800m$ and $600m$ the slope was initially covered by a debris blanket of triangular shape. This debris blanket had a thickness varying from a few centimeters at the top of the triangle to ten meters at its basis. On its lower side there were deposits constituted by finer materials, which probably came from previous landslides. The blanket lay on a natural platform, the Maan Pin platform, constituted by shale and dolomite strata. That platform was about $200m$ long, and it was covered by a thick $10m$ clay layer. The platform was weakly dipped downwards; at its end began a steeper slope, with a difference of level of about $40m$. After this jump there was a long channel inclined on average of $19°$. On the right side of the platform, a hill of finer material was also interested by the developing flow. The reported landslide was probably triggered by the heavy rainstorm occurred on the preceding days. The water, flowing into the rock fractures, saturated the rock mass and therefore caused an overpressure under the clay layer. The overpressure decreased the effective stresses and led to the slide movement.

It is possible to divide the landslide in two phases. The first one is the slide of the clay layer and of the debris blanket which lay on the platform. After running out of the Maan Pin, this material fell into the $40m$ deep cross channel and decomposed into debris flow. The second phase of the landslide is a proper debris flow, which split in two branches. The main one ran downward along the gully and accumulated in a fan at the slope toe. The mass of the other branch eroded the hill, ran along a right gully and accumulated at the gully curve.

The total volume of the material involved in the landslide was estimated to be $1.000.000\ m^3$, about ten times larger than that of the initially detaching mass. This fact indicates that, during its flow, the debris both incorporated other material along the path and increased its voids ratio, thus increasing its total volume.

Our numerical simulation treated the entire process as a debris flow. This has been necessary in order to overcome the difficulty of establishing accurately the shape, the quantity and the velocity of the material at the passing from slide to debris flow.

**Fig. 6.** Topography of the Xi Kou landslide, as obtained from Jiang Yun et al. (1991). Bold lines represent the isolines acquired from the original map; the reconstructed isolines and the contours of the reported event zones are shown in the upper projection.

The topography of the site was reconstructed by acquiring the planimetry data reported by Jiang Yun et al. (1991). A total of 546 coordinate points on the map was tracked; an interpolating surface (Fig. 6) was then computed in order to obtain the topographical surface heights at regularly spaced lattice positions. An interpolation library routine based on a radial functions fit (PV-WAVE, 1993, routine RADBF) was employed. The resulting lattice was a $61 \times 151$ rectangular grid of $10m$ spaced points. The thickness and the composition of the debris layer were defined in every lattice cell according to the description of the site given by Jiang Yun et al. (1991). Three zones of the site were filled with mobilizable material, with the different compositions reported in Table 3. The water content was determined by assuming that, initially, the material was completely saturated: its entire void volume, which is found multiplying the solid volume by $e_0$, was assumed to be filled by water. The "in situ void ratio" $e_0 < e_{max}$ was determined from empiric tables (Lancellotta, 1987).

The results of the numerical simulation are depicted in Figs. 7–9. These figures show the debris height and the flow velocity vectors in the cells at different times, plotted over the planimetry of the site.

**Table 3.** Initial distribution of mobilizable material for the Xi Kou slide simulation

|  | Triangle | Platform | Hill |
|---|---|---|---|
| Shape | pyramidal | $10m$ layer | $14m$ conoid |
| $s_1$ % | 33.1 | 35.7 | 24.3 |
| $s_2$ % | 49.7 | 23.8 | 36.5 |
| $s_3$ % | 13.2 | 0 | 0 |
| $w$ % | 33.7 | 40.5 | 39.2 |

**Fig. 7.** Numerical simulation of the Xi Kou landslide: debris accumulation at different times and computed instantaneous flow rates (arrows). For clarity only rates exceeding $0.1 * q_{max}$ have been plotted. The contours of the reported slide are superimposed to the graph for reference.

**Fig. 8.** Numerical simulation of the Xi Kou landslide: water fraction (a) and class 3 fraction (b) in the body of the slide at $t = 80$, together with flow rates (arrows). For clarity only rates exceeding $0.1 * q_{max}$ have been plotted.

The first panel of Fig. 7 shows the situation before the landslide. The debris is mainly located in the upper part of the slope and covers the platform. The following panels of Fig. 7 report intermediate flow snapshots. The simulated movement develops within the boundaries of the real event, branching to the secondary direction and interesting the hill as reported. The computed flow velocities are consistent with the preferential flow paths that result from the local morphology.

It may also be seen (Fig. 8) how, at intermediate times, the model reproduces a characteristic front surge shape, with higher concentrations of the coarser material found at the boundaries of the active region.

The larger discrepancies between the simulation and the event reported by Jiang Yun et al. (1991) are encountered at the slope toe and along the secondary tongue. We note that they occur where the lack of data has forced us to resort to conjectures: at the toe, the concave shape of the valley is generated by the interpolation, and is due to the lack of surrounding isolines; for the hill, the choice of modelling it as a cone of finer materials may have been arbitrary. We think however that the model satisfactorily captures the features of the event. A better calibration would be possible with more accurate and extensive data about the granulometric composition of the flowing mass. A complete calibration of the model would also require a statistical exploration of the parameters space for the maximum likelihood of the simulation.

**Fig. 9.** Numerical hydrograph of the debris flow entering cell $(25, 60)$: flow rate/base area and angle with the y-axis versus time.

In the present case, the parameter values have been obtained empirically while trying to reproduce as better as possible the shape and the extension of the area invaded by the debris flow and the event duration. These values are reported in Table 4. The simplest square lattice connectivity was used here.

## 5  Statistical properties

The debris flow we model is essentially a dissipative process. Even though our model does not account for an energy balance, the global dissipative character is apparent when evaluating the potential energy of the material distribution, which is

$$W(t) = gS \sum_i \rho(i,t) h(i,t) \left( z(i) + \frac{h(i,t)}{2} \right). \qquad (27)$$

As can be seen from Fig. 10, $W(t)$ is asymptotically decreasing, with small and irregularly spaced bursts of activity. These can be interpreted as temporary fluctuations, caused by the advection effect, which can occasionally draw some material counter-slope. Changes in the slope of $W(t)$ are connected with the development of the different phases of the landslide, and are related to the actual debris distribution over the topography.

**Table 4.** Parameter values for the Xi Kou slide simulation

| $Q_{-1}$ | $Q_0$ | $Q_1$ | $Q_2$ | $Q_3$ | $\alpha$ | $\beta$ | $\Delta t$ |
|---|---|---|---|---|---|---|---|
| 0.030 | 0.030 | 0.013 | 0.020 | 0.040 | 0.01 | 0.03 | 0.02 |

**Fig. 10.** Potential energy of the debris layer in time. The inset shows a zoom-in of the final part of the curve.

Following the line of investigation of the sandpile literature, we looked for a distinctive nonlinear behavior of the system by examining some of its statistical properties. In order to determine if the observed behavior could be described properly as intermittent, as happens for other nonlinear systems, we followed the procedure outlined by Provenzale et al. (1993). This requires a multifractal analysis of some signal that accounts for the global behavior of the system in time. We considered, as representative of the whole process, the cumulative height $M(t)$ of the material that is moved at each time step, i.e.,

$$M(t) = \sum_i \sum_{v=1}^b [q_{-1}(i,v,t) + (1+\alpha)\sum_{k=0}^3 q_k(i,v,t)]. \quad (28)$$

We presume that this quantity may be some sort of analogous of the avalanche size for the sandpile models. We decided to refer to time series generated by the Xi Kou simulation, although the procedure might have been applied as well to a completely idealized event. The time series of $M(t)$, plotted both in lin-lin and in log-log scale is shown in Fig. 11. Both a power law trend, which is the effect of the asymptotic behavior mentioned in Sect. 2.2.2, and the presence of irregular bursts are apparent in this plots.

We analyzed two sub-series consisting of 32768 points of this signal, starting from $t = 1000$ and from $t = 1844.64$. We estimated that, at these times, the most of the mass movement had already taken part (compare, e.g., the last two panels of Fig. 7), and the residual process could be considered nearly stationary. This is not entirely true, since the statistical properties of the two series are not found to be identical, and since a background trend is still present. A detailed analysis shows, however, that this trend does not affect the following conclusions.

We then considered the numerical derivative $D(t) = [M(t + \Delta t) - M(t)]/\Delta t$, i.e., a fluctuation signal. This numerical differentiation is sufficient for filtering out the small and much slower relaxation trend present in the two sub-series.

**Fig. 11.** Cumulative movement of material across the lattice during time, $M(t)$. Insets (a) and (b) show details of the curve; inset (c) is a log-log plot of the same. Y-axis units are of h/time.

Fourier analysis of $D(t)$ shows a noisy spectrum (Fig. 12), without the clean power law decay exhibited by simpler sandpile models for the distribution of the avalanche sizes.

Figure 13 shows the cumulative histogram of the fluctuations, which evidently follow a non-gaussian statistic. The values of the first two moments of the distributions and some other fine details slightly differ for the two sub-series. However, both histograms have a shape which is much closer to that of exponential distributions, with some extra features like a secondary bump for negative deviations.

In parallel, we created two surrogate time series by phase randomization of the amplitude spectra of the original ones. The probability distributions of their fluctuations (Fig. 14), this time strikingly gaussian, denote that the nontrivial statistic is reflected in a phase locking of the Fourier components. One conclusion of this fact is that the instantaneous transfers across the lat-tice cannot be uncorrelated, and the landslide process has memory of its history.

**Fig. 12.** Amplitude spectra of $D(t)$. Dotted lines are a power law fit of the last decade of the spectrum, included for reference. Here and afterwards (a) and (b) refer to the two sub-series.

**Fig. 13.** Histograms of the fluctuations of the two sub-series of $M(t)$. The dotted parabolas correspond to gaussian distributions with the same first two moments.

**Fig. 15.** Spectra of $d_q$ for the sub-series of M(t) (□) and for surrogates (◦). Error bars represent 95% confidence intervals on the line fit.

To characterize the intermittency we then computed the generalized fractal dimension $d_q$ of the sub-series, along with that of their surrogates. Following Provenzale et al. (1993) and the references cited therein, we first computed the partition function

$$B_\delta(\epsilon, q) = \sum_i \left( \int_{t_i-\epsilon/2}^{t_i+\epsilon/2} \delta(t) dt \right)^q . \qquad (29)$$

In the latter equation, $\delta(t)$ is an appropriate measure function of the activity of the signal; numerical inspection proved $D(t)^2$ to be a suitable one. If the signal is multifractal, then, for small $\epsilon$, $B_\delta(\epsilon, q)$ scales like

$$B_\delta(\epsilon, q) \propto \epsilon^{(q-1)d_q} . \qquad (30)$$

The generalized dimension $d_q$ is thus found for every $q$ by a linear fit of $B_\delta(\epsilon, q)$ in log-log coordinates, within the scaling range. The computation shows a good power law scaling of $B_\delta(\epsilon, q)$ on the range $\Delta t$ and $T/30$, that is over more than 3 decades. The multifractal spectra $d_q$ in the range $0 < q < 4$ are shown in Fig. 15. The surrogates have higher values of $d_q$ than those of the original series, and also a slower decrease of $d_q$ with increasing $q$. We therefore conclude that the original series have a self-affine behavior, which is peculiar since it is lost after the phase randomization; our signals $M(t)$ can be then legitimately considered as multifractal.

## 6 Conclusions and perspectives

We have set up a model of debris flow which is schematic and therefore computationally light. A preliminary validation, based on the field data that was presently available to us, showed that the model is able to capture most of the features of the natural process. In light of this agreement, it would be interesting to perfection the model and to calibrate it with more complete data, thus establishing a correspondence between its parameters and field measurements of geotechnical significance. A complete calibration shall base itself on a fully monitored and accessible landslide event, in which the granulometric composition at various locations and the extent of the slide are exactly known. A refinement of the flow rules may also be required.

We have seen that the model behaves globally as a nonlinear system, giving rise to intermittent signals. In this our model is probably just an elaboration of the basic sandpile models, and follows their statistical characteristics.

It would be interesting, but it falls beyond the scope of this paper, to make an explicit connection between our model and other lattice methods. The construction of our model was indeed suggested by the extensive examples of lattice gases, but many important differences prevent us from making use, at this stage, of known results for the latter. Some analogies can be drawn, and we just want to point at them, without attempting to establish the complete reasoning that connects lattice gases with continuum equations (Frisch et al., 1986).

In lattice gases, self-excluding particles travel along the lattice branches, and collision rules that conserve mass, momentum and energy are implemented. Such particles may have different masses, or be of different species. In our model we have different moving materials, but only material mass is conserved. The role of critical directions is perhaps equated to that of the gas molecules, in the sense that momentum is extracted from the cell in those directions. Our model may be perhaps compared in this context with lattice Boltzmann automata, because it employs real state variables rather than discrete ones. This is also suggested by the fact that our automaton relies on a sort of local equilibrium assumption, since models a non-stationary process by assuming the local validity of an equilibrium flow law. Likewise, Boltzmann automata use equilibrium probability distribution functions at each lattice site, and can already work with a single relaxation time form of the collision operator (Chen et al., 1992).

**Fig. 14.** Histograms of the fluctuations of the surrogate signals. The format is the same as in Fig. 13.

Another parallel that could be made would be viewing our landslide automaton as a sort of percolation model (see, e.g., Sahimi, 1993, for a review), in which the percolation network is continuously modified by the advected material.

In our primary point of view, however, just a descriptive model was sought. In this aim, a finer resolution of the studied domain is expected to lead just to a greater accuracy of the model. In a lattice gas, instead, a finer resolution is considered in order to perform macro averages, which allow to recover continuum equations in the asymptotic limit. That might also be done with our set of cell rules, which are more complicated than the ones of the Navier-Stokes automata, perhaps in the direction of recovering some non-Newtonian debris flow equation. However, we note that even for the empiric constitutive relations there is no complete agreement in the literature, where generally just particular regimes of equilibrium flows are considered. In the general case, for flows with a free surface, varying slope, flow depth and flow rate, vertically averaged De Saint-Venant type equations are usually found. At the present state, we would not know to which continuum equation the model should tend in a thermodynamic limit. Moreover, our interest was in the large scale modelling of a landslide site, rather than on the fundamental form of the constitutive debris flow equations.

Once better founded, we believe that this model could be not only a tool for a better simulation of hazardous events, but also open the possibility of numerical investigation on more abstract topics, such as the connection between the statistical properties of the topography of arbitrary landscapes and that of the landslides taking part on them, or the response of unstable debris accumulations to random perturbations.

*Acknowledgements.* We are very grateful to Prof. G. P. Giani, who introduced us to the problem and supported this study. We also thank Drs. G. Boffetta and A. Provenzale for providing many helpful and encouraging comments. This work has been carried on by means of the computer facilities at the Istituto di Cosmogefisica of the CNR, Torino, for whose warm hospitality we are grateful. E. S. wishes also to thank C. Siciliano for having provided the preprint of Digregorio et al. (1994), and Prof. W. Kantor for advice in revising the manuscript.